\documentstyle[12pt]{article} 
\normalbaselines

\font\frak=eufm10 scaled\magstep1

\font\black=msbm10 scaled\magstep1
 2

\def\goth #1{\hbox{{\frak #1}}}

\def\Bbb #1{\hbox{{\black #1}}}
\def\v #1{\vert #1\vert}             
 
\def\m #1 #2{(-1)^{{\v #1} {\v #2}}} 
\def\pd#1#2{\frac{\partial#1}{\partial#2}}
\def\pois#1#2{\{#1,#2\}}            
\def\<#1>{\langle#1\rangle}        
\def\>#1{{\bf #1}}                
\def\f(#1,#2){\frac{#1}{#2}}

\def\dt2#1{\frac{d^2 #1}{dt^2}}

\def\a{\alpha}                    
\def\ad{\opname{ad}}              

\def\opname#1{\mathop{\rm#1}\nolimits} 

\def\bbuildrel#1_#2^#3{\mathrel{
     \mathop{\kern 1pt#1}\limits_{#2}^{#3}}}

\def\frac#1#2{{#1\over  #2}}      
\def\Ga{\Gamma}                   
\def\ga{\gamma}                   
\def\harr#1#2{\smash{\mathop{\hbox to .5in{\rightarrowfill}}
     \limits^{\scriptstyle#1}_{\scriptstyle#2}}}
\def\hharr#1#2{\smash{\mathop{\hbox to .3in{\rightarrowfill}}
     \limits^{\scriptstyle#1}_{\scriptstyle#2}}}
\def\harrt#1#2{\smash{\mathop{\overline{\hbox to .5in{\hrulefill}}}
     \limits^{\scriptstyle#1}_{\scriptstyle#2}}}
\def\ID{\relax{1\kern-.24 em\rm l}}  
\def\LG{{\goth g}}                
\def\ker{\opname{ker}}            
\def\la{\lambda}                  
\def\o{\omega}                    
\def\o+{\oplus}                   
\def\O+{\bigoplus}                
\def\pd#1#2{{{\partial#1}\over{\partial#2}}}  
\def\R{{\Bbb R}}                  

\def\v{\vee}                      

\def\w{\omega}                    
\def\X{{\goth X}}                 
\def\1{\'{\i}}                    
\def\3{\sharp}                    
\def\7{\dagger}                   
\def\.{\cdot}                     
\def\:{\colon}                    
\def\<#1>{\langle#1\rangle}       
\def\hw{\hat\omega}
\def\pois#1#2{\{#1,#2\}}            

\def\hw{\hat\omega}
\def\BL{{\Bbb L}}                   
\def\bea{\begin{eqnarray}}
\def\eea{\end{eqnarray}}
\def\Aad{\hbox{{\bf\rm Ad}}}
\def\aad{\hbox{{\bf\rm ad}}}
\begin{document}
\bibliographystyle{unsrt}

\vbox {\vspace{6mm}} 

\begin{center}
{\large \bf  SYMPLECTIC AND LIE ALGEBRAIC TECHNIQUES
 \\[2mm]
 IN GEOMETRIC OPTICS}\\[7mm]
J.F. Cari\~nena${}^{\dag}$, C. L\'opez${}^{\ddag}$
 and J. Nasarre*\\
{\it ${}^{\dag}$Departamento de F\'{\i}sica Te\'orica,  Universidad de Zaragoza, Zaragoza 50.009\\
 ${}^{\ddag}$ Departamento de Matem\'atica Aplicada, CPSI,
Universidad de Zaragoza, Zaragoza 50.015}
\\{\it *IES Miguel Catal\'an},
{\it Isabel la Cat\'olica 3,  Zaragoza 50.009}\\
\end{center}

\vspace{2mm}

\begin{abstract}

We will show the usefulness of the tools of Symplectic and
Presymplectic Geometry and the corresponding Lie algebraic methods in
different problems in Geometric Optics.

\end{abstract}

\section{Introduction: Symplectic and Presymplectic geometry}

Geometric techniques have been applied to physics for more than 50 years in
many different
ways and they have provided powerful methods of dealing with classical
problems from a new geometric perspective. Linear representations of groups,
vector fields, forms,
exterior differential calculus, Lie groups, fibre bundles,
connections and Riemannian Geometry,
symmetry and reduction of differential equations, etc...,
are now well established tools in modern physics. Now, after more than
twenty years of using Lie algebraic mehods in Optics by Dragt, Forest,
Sternberg,
Wolf and their coworkers, we aim here
to establish the appropriate
geometric setting for Geometric Optics. Applications in computation of
aberrations for different orders will also be pointed out.

The basic geometric structure for the description of classical
(and even quantum) systems is that of symplectic manifold.
A   {\it  symplectic manifold\/} is a pair
$(M,\w)$ where  $\w$ is a nondegenerated closed  2-form in
 $M$.  If $\w$ is exact we will say  that $(M,\w)$ is
 an {\it exact  symplectic manifold\/}. Let  $\hw :\X (M)\to  \bigwedge^1(M)$ be
 given by $\hw(X) =
i(X)\,\w, \hw(X) Y=\w(X,Y)$. The two--form
$\w$ is said to be nondegenerate when  $\hw $ is a bijective map.
Then $M$ is evendimensional and it may be used to identify vector fields
on $M$ with 1--forms on $M$. Vector fields $X_H$ corresponding to
exact 1--forms $dH$ are called Hamiltonian vector fields. The 2--form
$\w$ is said to be closed if $d\w=0$.

The simplest example is
${\Bbb R}^{2n}$ with coordinates
$(q^1,\ldots ,q^n,p_1,\ldots, p_n)$ endowed with the constant 2-form
$\w=\sum_{i=1}^ndq^i\wedge dp_i$. Closedness of $\w$ is very important because
Darboux theorem establishes that for any point   $u \in M$ there exists a local
chart
 $(U,\phi)$ such that if  $\phi = (q^1,\dots,q^n;
p_1,\dots,p_n)$, then $\w|_U = \sum_{i=1}^n dq^i \wedge dp_i$.
Consequently, the example above is the local prototype of a symplectic
manifold.It is also well known that if $Q$ is the configuration space of a
system, its cotangent bundle, $T^*Q=\bigcup_{q\in Q}T^*_qQ$, called phase space,
is endowed with
a canonical 1--form $\theta$ on $T^*Q$ such that $(T^*Q,-d\theta)$ is an
exact symplectic manifold. More especifically,
if $(q^1,\ldots,q^n)$ are coordinates in $Q$ then $(q^1,\ldots,q^n,p_1,
\ldots,p_n)$ are coordinates in $T^*Q$ and $\theta=\sum_{i=1}^n p_i\,
dq^i$,
$ \w=\sum_{i=1}^n dq^i \wedge dp_i$.

A Hamiltonian dynamical systems is a  triplet
$(M,\omega,H)$ where $M$ is a differentiable manifold,
$\omega\in Z^2(M)$ is a symplectic form in $M$ and
$H\in C^\infty (M)$ is a function called Hamiltonian. The dynamical vector field
$X_H$ is then the solution of the equation $i(X_H)\omega=dH$. In
the example above
$$ X_H = \pd H{p_i} \pd{}{q^i} - \pd H{q^i} \pd{}{p_i}.$$
and the same expression is valid in Darboux coordinates for any
Hamiltonian dynamical system.A Poisson bracket can be introduced in a symplectic
manifold $(M,\w)$ by $\pois FG=X_GF=\w(X_F,X_G)$.
Then, closedness of $\w$ is equivalent to Jacobi identity for P.B.
Moreover, it may be shown that
$\sigma=\hw^{-1}\circ d: C^{\infty}(M)\to \X(M)$ is a Lie algebra
 homomorphism.

 A presymplectic manifold is a pair $(M,\w)$ such that
$M$ is a differentiable manifold and $\w$ is  a constant rank  closed 2--form in
$M$. The kernel of $\w$ defines an integrable  distribution (because $d\w=0$)
and when the set of leaves is a manifold, it can be endowed with a symplectic
structure. This process is called  reduction of the presymplectic structure.

Very interesting  examples of HDS are those defined by regular
 Lagrangians, $(TQ, \omega_L, E_L)$,
with $\omega_L=-d\theta_L=-d(dL\circ S)$, $E_L=\Delta L-L$.
More accurately, the geometric approach to the Lagrangian description
makes use of the geometry of the tangent bundle of the
configuration space that we will shortly review. The  tangent bundle $\tau_Q\:
TQ\to Q$ is characterized by the existence of a vector field generating
dilations  along the fibres, called Liouville vector field,
  ${\Delta}  {\in }{ \goth X}(TQ)$,
  and
the vertical endomorphism which is a $ (1,1)$--tensor field $S$ in $TQ$
that  in a natural coordinate system for $TQ$, induced from a chart
in $Q$, are
$ \Delta =v^i\partial/\partial v^i$,
and  $ S= (\partial/\partial v^i) \otimes dq{^i}.$ Given a function $L\in
C^\infty (TQ)$,
 we  define the 1--form  $\theta _L\in \bigwedge ^1(TQ)$ by
$\theta _L=dL\circ S$. When the exact 2--form $\omega _L= -d\theta _L$
 is nondegenerate the
 Lagrangain $L$ is called regular and then $(TQ,\omega_L)$ is a symplectic
 manifold. The energy function $E_L$ is given by $ E_L={\Delta  }(L) - L$.
The
 coordinate  expressions are
$ {\theta}_L= (\partial L/\partial v^i) dq{^i}$ and $ E_L=v^i(\partial
L/\partial v^i)
 - L $.

 \section{Symplectic structures in Geometric Optics}

The set
 of oriented
geodesics of a Riemannian manifold can be endowed with
 a symplectic structure and in particular the set of
oriented straightlines in the plane, which is   the set
of light rays  in a two--dimensional constant rank medium,
can be endowed with a symplectic structure. Moreover,
it can be considered as the
cotangent bundle of the one--dimensional sphere $S^1$.
If an origin $O$
has been chosen in the plane,
 every oriented straightline  that does not pass through the point $O$
is characterized by  a unit vector $\bf s$ pointing in the line
direction and a vector $\bf v$ orthogonal to  $\bf s$  with end on the line and
 origin in $O$. The straightlines of a pencil of oriented parallel lines
  are characterized by proportional vectors $\bf v$ and
 the same $\bf s$.
Straightlines passing through $O$ with direction given by
 $\bf s$
correspond to $\bf v=0$.
The vectors $\bf v$ and
$\bf s$ being orthogonal and ${\bf s}\cdot {\bf s}=1$, the couple
$({\bf s},{\bf v})$ can be seen as a tangent vector to the
unit circle $S^1$ at the point described by $\bf s$.

The Riemannian metric in $S^1$ can be used to identify in each point $\bf s$
the tangent space $T_{\bf s}S^1$ with its dual space
$T^*_{\bf s}S^1$ and therefore   the tangent bundle $TS^1$
with the cotangent bundle $T^*S^1$. This identification shows us
that the space of oriented straightlines in the Euclidean two--dimensional
space can be  endowed with an exact symplectic structure which corresponds
to the canonical   structure for the cotangent bundle $T^*S^1$. The study of oriented
straightlines in Euclidean
three--dimensional space follows a
similar pattern.

A choice of coordinates in the base space will provide us Darboux
coordinates: a good
choice will be an angle coordinate.
A straightline
$y=m\, x+ b$  with slope $m= \tan \theta$ will be represented
by a vector orthogonal to the vector ${\bf s}=(\cos \theta,
\sin\theta)$, and length $b\, \cos\theta$,   namely,
${\bf v}= b \cos \theta\, \pd {}{\theta}$. The vector $\partial/\partial\theta$ is
unitary  in the Euclidean metric, and then the point
$(\theta,p_\theta)\in T^*S^1$
corresponding to   $(\theta, v_\theta)$ is   $p_\theta=v_\theta$.
The symplectic form in $TS^1$ translated from the canonical
symplectic structure in $T^*S^1$ $\omega_0=d\theta\wedge dp_\theta$ will be
$\w= d\theta\wedge d(b\, \cos \theta)=d(\sin \theta)\wedge db$. Therefore,
Darboux coordinates for $\w$ adapted to the cotangent  structure are not only
$(\theta, b\cos\theta)$ but also   $q= \sin \theta$, $ p=b$,
which are more appropriate  from the experimental viewpoint. So, the flat
screens
arise here as a good choice for Darboux coordinates.

  The choice usually done in   Geometric Optics
is ${\bf s}\cdot{\bf s}=n^2$, the Darboux coordinate $q$ then being
 $q=n\, \sin\theta$. This leads to the image of the Descartes sphere,
a sphere of radius
 $n$ whose  points describe the ray directions.  In the more general case of a variable refractive index, we
recall that light rays  trajectories in Geometric Optics are determined by
Fermat's principle: the ray path connecting two points is
the one making stationary the optical length:
$\delta \int_\ga  n\, ds=0$.

 This corresponds to
the well--known Hamilton's principle of Classical Mechanics with an \lq\lq
optical  Lagrangian" $L=n\, \sqrt{v_x^2+v_y^2+v_z^2}$, which is a
differentiable
function in $T{\Bbb R}^3$ up to the zero section. In other words,
the  mechanical problem corresponding to     Fermat's principle
leads to a singular Lagrangian $L(q,v)=[g(v,v)]^{1/2}$, where $g$ is
 a metric conformal to the Euclidean metric $g_0$,
$g(v,w)= n^2 g_0(v,w). $ $L$ is an
homogeneous function of degree one in the velocities and consequently $L$
 is singular and the corresponding energy function vanishes identically.

It was shown in \cite{CL91} that it is possible
to relate the solutions of the Euler--Lagrange equations for $L$ with
those of
the regular Lagrangian $\BL=\frac 12 L^2$, up to a reparametrization.
$\Bbb L$ is quadratic in velocities and   the solution
  $\Ga_{\Bbb L}$ of the  equation
$i(\Ga_{\Bbb L})\w_{\Bbb L}=
dE_{\Bbb L}=d{\Bbb L}$ is not only a second order differential equation
vector field
 but also a spray,
 the projection onto
 ${\Bbb R}^3$ of its integral curves being the geodesics of the Levi--Civita
connection defined by $g$. The kernel of $\w_L$ is two--dimensional
and it is generated by  $\Ga_{\Bbb L}$  and the Liouville vector field $\Delta$.
The distribution $\ker \w_L$ is integrable because
$d\w_L=0$;
the distribution is also
generated by $\Delta$ and
 $K=\frac 1{v^3}  \Ga_{\Bbb L}$, for which $[\Delta, K]=0$.

If the refractive index
for an optical system
 depends only on $x^3$ and
 the region in which the index is not constant is bounded, we can choose
Darboux coordinates by fixing a $x^3$ outside this region and taking Darboux
coordinates for the corresponding problem of constant index \cite{CN96a}. This justify
the choice of coordinates for the  ingoing and outgoing light
rays in the constant index media, i.e. it shows the convenience of
using  flat screens in far enough regions on the left and right respectively,
and then this  change of Darboux
coordinates seems to be, from an active viewpoint,
a canonical transformation. Similar results can be obtained  (see \cite{CN96b})
 for nonisotropic media
 for  which the refractive index
depends only on the ray direction, i.e. $n=n(v)$
and $\Delta n=0$. The only difference is that
$\w_{\Bbb L}$ may be singular, but in the regular
 case all works properly.

\section{Group theoretical approximations}

 Mathematical expressions like $x'=f(x)$ admit
two different interpretations. In the {\sl alias\/} interpretation
$x$ and $x'$ are coordinates of the same point in two different
coordinate systems, while in the  {\sl ad libi\/} interpretation
$x$ are the coordinates of a point and $x'$ those of its image
under the transformation defined by $f$. In this sense a
change of Darboux cordinates
can be seen as a canonical transformation in $\R^{2n}$, and in
particular, when Darboux coordinates are chosen as indicated above,
the passage of the set of light-rays through an optical device can be
considered as a canonical transformation. Moreover, we can split an optical
system in two subsystems and the canonical transformation factorizes as a
product of two canonical transformations. Even if the group of canonical
transformations is not a Lie group, any element $g$ can
be written as the exponential of an element in its Lie algebra, the set of
Hamiltonian systems, $g=\exp X_f$. Symmetry of the optical system leads  to
reduction, and then to a lower number of degrees of freedom.

There exist formulae generalizing Baker-Campbell-Hausdorff for composition of
generating functions, both in an exact or approximate way. Most of
approximation formulae substitute the generating functions by
a power series development and then only keep some terms, giving rise
in this way to aberrations. For instance, if we only consider quadratic terms,
we will get the linear approximation.

The fundamental algebraic  ingredients for the theory of approximate groups are
the concepts of enveloping algebra ${\cal U}$ and symmetric algebra $\Sigma$ of
a Lie algebra ${\goth g}$. Essentially, if $\{X_1,\ldots,X_n\}$ is a basis of
 ${\goth g}$, then  $\Sigma$ is the algebra of polynomials in  $\{X_1,\ldots,
X_n\}$. Both   ${\cal U}$ and $\Sigma$ have graded Lie algebra structures
extending that of ${\goth g}$, which can be identified as a subalgebra of ${\cal
U}$ and $\Sigma$. In the same way as  ${\goth g}$ can be seen as the set of
linear functions on
${\goth g}^*$, the symmetric algebra can be considered as the set
of polynomials on ${\goth g}^*$- The adjoint action of $G$ on ${\goth g}$
can be extended to an action $\Aad: G \times \Sigma \to      \Sigma$ in such
way that
$\Aad (g)$ is linear for each $ g\in G$ and
$ \Aad (g)  (p_1.p_2) = \Aad (g) (p_1) . \Aad (g)(p_2) $.
The extension of the adjoint action of the Lie algebra,
$\ad:   \LG  \times   \LG  \to      \LG,\quad
\ad(a,b)=\ad(a)(b) =[a,b]$, is the adjoint action of the
symmetric  algebra $\Sigma$:
	$$\aad: \Sigma \times  \Sigma \to    \Sigma ,\quad
     \aad (p_1,p_2) = \aad (p_1)(p_2) =
 [p_1,p_2]_\Sigma.$$

	For any $p\in       \Sigma$, we can also  consider the
formal  transformation of $\Sigma$,
	$\phi (p) : \Sigma\to     \Sigma$,
	$$p'  \to
\phi(p)(p')=\exp (\aad (p)) (p') =
 p' + [p,p']_\Sigma+  1/2 [p,[p,p']_\Sigma]_\Sigma
 +\ldots ,$$
 and we should now consider the elements $p$ for which such expresion is
meaningful. They span a group $G_{\Sigma}$. The enlarged action of it reduces to
the identity when acting on the set $\Sigma^I=\{ p\in       \Sigma \mid
 [a,p]_\Sigma=0,\, \forall   a\in       \LG \}$
of the polynomial Casimir elements  of
 $\LG$. We shall then pass to the quotient graded
Lie algebra $\Sigma^C=\Sigma/\Sigma^I$. Finally for approximation we will
consider for each $r\in {\Bbb N}$ the ideal spanned by
$H_r=\bigoplus_{t>r}\Sigma_t$, and then $\phi([p])H_r\subset H_r$, and therefore
it induces a map $\Phi^r([p])$.

\section{ Perturbative treatment of aberrating optical
systems using Weyl group}

A model for geometrical optics in a plane is obtained from the Weyl group
$W(1)$: it  is a  three--dimensional Lie group,
 with elements $g\in   W(1)$ labelled by $  g=(\mu_1 ; \nu)$,
$\mu_1\in \R^2$, $\nu\in       \R$,
and composition law
$g'g=(\mu_1+\mu'_1,\nu'+\nu+\frac 12
\mu'_1\wedge \mu_1)
$
where $\wedge$  denotes $(a,b)\wedge (c,d)=ad- bc$.

	A basis for the Lie algebra ${\goth w}(1)$,  is given by
$ Q= \partial _{a} - \frac 12 b\,\partial _\nu,\,
P= \partial _{b} - \frac 12 a\,\partial _\nu,\,
I=\partial _\nu$ with Lie brackets $  [Q,I]= [P,I]=0,\,
	[Q,P]= I.$		An infinite-dimensional basis for
the associated symmetric algebra $\Sigma$ is given by $\{1,I,Q,P,I^2,$
$ IQ,IP,Q^2,QP,P^2,\ldots \}$.
$\phi (\la_1I+\la_2Q+\la_3P)$ is in fact an
element of $G$.
Its generalized adjoint action on $\Sigma$ preserves each subspace
$\Sigma_r$. It is enough to know its
action on $\Sigma_1$, $(c_1I + c_2Q + c_3P \rightarrow
((c_1 -\la_3 c_2 + \la_2 c_3)I + c_2Q + c_3P)$.
Another typical element of $\,G_\Sigma$ is given by
$\phi (\nu _1I^2+\nu _2IQ+\nu _3IP+\nu _4Q^2+\nu _5QP+\nu _6P^2)$,
 which in fact is only a
 formal map.
 However, its projected maps on $\Bbb P _{\rm r}$ are well
defined. The
infinitesimal adjoint action $\aad(\nu _1I^2+\nu _2IQ+\nu _3IP+\nu _4Q^2
+\nu _5QP+\nu _6P^2)$ maps
 each $\Sigma _r$ onto $\Sigma _{r+1}$, so that only its action on $\Sigma _1$
 is not trivial when considering
its projected map on $\Bbb P _{\rm 2}$=$\Sigma_0\oplus \Sigma_1\oplus \Sigma_2$.

The Casimir elements of ${\goth w}(1)$ are the
polynomial functions on $I $ and 1.  A basis for the quotient algebra
$\Sigma^C$ is $\{[1]^{\rm c}, [Q]^{\rm c}, [P]^{\rm c},  [Q^2]^{\rm c},
[QP]^{\rm c}, [P^2]^{\rm c}, \ldots \} $.  The reduction
process can be obtained by quotient by the ideal generated by the Casimir $I$. A typical
element of the group $G_{\Sigma ^C}$ is $\phi ^C(\nu _1[Q^2]^c+\nu _2[QP]^c+\nu
_3[P^2]^c)$. Its projected action on $\Bbb P ^C_{\rm 2}$ is
given by the matrix
$$\phi ^C(\nu _1[Q^2]^c+\nu _2[QP]^c+\nu _3[P^2]^c)=\left(\matrix{1&0_{1\times
2}& 0_{1\times 3}\cr0_{2\times 1}&M&0_{2\times 3}\cr
 0_{3\times 1}&0_{3\times 2}&D^2(M)\cr}\right)
$$
where $M\in SL(2,\R)$ and $D^2(M) $ is the image of $M$ in the
 three-dimensional representation:
$$M=\left(\matrix{ \a&\beta\cr \gamma&\delta\cr}\right),
\qquad D^2(M)=\left(\matrix{\a^2&\a\beta&\beta^2\cr 2\a \gamma &\a\delta+\beta
\gamma &2\beta\delta\cr
 \gamma^2& \gamma\delta&\delta^2\cr}\right).
$$

The matrix $M$
 associated to the map $\phi ^C(\nu _1[Q^2]^c+\nu_2[QP]^c+\nu_3[P^2]^c)$
 is given by
$$M=\left(\matrix{ \cosh \w-\frac {\nu_2}\w\sinh \w&2\frac {\nu_1}\w\sinh \w\cr
-2\frac {\nu_3}\w\sinh \w&\cosh \w+\frac {\nu_2}\w\sinh \w\cr}\right),
 \ {\rm{with}}\ \w=\pm \sqrt {\nu_2^2-4\nu_1\nu_3}.
$$

A  matrix representation for the group $G_{\Sigma_r^c}$
provides  a perturbative treatment of $(r-1)$-th order
(in 1+1  dimensions). Using the factorization
 theorem and the axial symmetry of the system, only transformations
 of type $\phi^C([p]^c) $ with $[p]^c\in    \Sigma^C_2$ or $\Sigma^C_4$ must be
considered for third order aberrations. $\phi^c(\nu
_1[Q^2]^c+\nu_2[QP]^c+\nu_3[P^2]^c)$ has a matrix representation immediate
generalization of the above mentioned on  $\Bbb P ^C_{\rm 2}$. The
element  $\phi^4 (\mu_1[Q^4]^c+\mu_2[Q^3P]^c+\ldots
 +\mu_5[P^4]^c)$ is
$$\phi^4
(\mu_1[Q^4]^c+\mu_2[Q^3P]^c+\ldots
 +\mu_5[P^4]^c)=\left(\matrix{  1&0&0&0&0\cr 0&I& 0&0&0\cr 0&0&I&0&0\cr
0&M_1&0&I&0\cr  0&0&M_2&0&I\cr}\right). $$
 $$M_1= \left(\matrix{-\mu_2&4\mu_1\cr -2\mu_3&3\mu_2\cr -3\mu_4&2\mu_3
\cr -4\mu_5&\mu_4\cr}\right), \quad M_2=\left( \matrix{ -2\mu_2&4\mu_1&0\cr
-4\mu_3&2\mu_2&8\mu_1\cr  -6\mu_4&0&6\mu_2\cr -8\mu_5&-2\mu_4&4\mu_3\cr
0&-4\mu_5&2\mu_4\cr}\right)$$
The representation splits into two,  acting
respectively  on the even and odd degree subspaces.
The fisrt one
can be used to find the composition law and the second one
can be
used to find the approximate coadjoint action on the coordinate  functions
$q$ and $p$.

The free  propagation,
till the  third degree approximation is $ p' = p,\,
	q' = q + \frac znp + \frac z{2n^3}p^3 $,
and the group element will be of the form  $\phi ([p_2]^c)\circ \phi ([p_4]^c)$,
with $[p_i]^c\in \Sigma_i^c$. The matrix $M\in  SL(2,\R)$ associated to
$\phi ([p_2]^c)$ is
$M=\left(\matrix{
1&0\cr  z/n&1\cr}\right) $ while $[p_4]^c= -\frac z{8n^3}[P^4]^c$,  so
that, $\{ \mu_1 ,\mu_2 ,\mu_3 ,\mu_4 ,\mu_5 \} =
\{ 0 ,0 ,0 ,0 , -\frac z{8n^3} \}$ determines $M_1$ and $M_2$.

The same
calculus for a refracting surface gives
$$M=\left(\matrix{ 1&\frac {n_1-n_2}R\cr 0&1\cr}\right), {\rm
{\ and\ }} { \mu } =
\{ \frac {3 n_2 - n_1 - 2 \frac {n_2^2}{n_1}}{8R^3} ,
\frac {1 - \frac {n_2}{n_1}}{2R^2} ,
\frac {\frac {1}{n_2} - \frac {1}{n_1}}{4R} ,0 ,0 \}.
$$

\section{Example}

In the design of a doublet we have seven of these basic systems concatenated
so that the total system (chosen to be telescopic) is obtained in third
order approximation by composition formulae of the corresponding
third order aberration group.
The composition of the systems $(M_2, { \mu_2})$ and $(M_1,{ \mu_1})$,
with $M_i$ the linear approximation matrices and ${ \mu_i}$ the coefficients
of the fourth order polynomials, is obtained by the formula
$(M_2 M_1, D^4(M_1^{-1}){ \mu_2} + { \mu_1})$, $D^4(M)$ being the former
representation of matrix $M$ on the fourth order polynomial space. In our
example, a concatenation of compositions for the doublet gives way to a total
linear approximation matrix  (on which we can impose the telescopic condition
$\gamma = 0$ and a given factor of magnification, say $\delta = 5$) and a total
fourth order polynomial. Fixing the refractive indexes of the lenses as
 $n_1 =
7/4$ and $n_2 = 9/4$ we obtain a seven parameter system of equations.
The polynomic expression of ${ \mu}$ in terms of $S_i$ and $z_j$ (the radii
and lengths of the lenses) is a set of eleven to sixteen
degree polynomials, so that numerical calculus should be used to obtain
solutions with zero third order aberrations.

The example does not try to be a realistic design, in which a four
dimensional space should be used
on the $q$'s and $p$'s, chromatic aberration should be taken into account
through the dependence on the refractive index with the wave length, and
stability of the solution to errors on the  parameters should be considered.
For our playing design system, and in order to simplify the calculus,
we can fix some of the parameters in terms of the other ones, say
$S_2=S_3=z_1/4$, $z_3=z_1$ and $z_2=2z_1$, so that only three parameters
are left free. Taking into account the telescopic condition and the given factor
of magnification we are left with just one parameter, which can be used to
minimize the square of the ${ \mu}$ vector.

A numerical solution obtained for this simplified case is
$z_1=0.5925$, which gives a linear approximations for the total system
and a  total vector ${ \mu}_{\rm{ total}}$
$$
M_{\rm{ total}}=\left(\matrix{ 0.2 & 1.7587 \cr
  2.98\,{{10}^{-19}} & 5. \cr}\right), \ { \mu}_{\rm{ total}}=\{ 0.1295,-1.4124,
  1.2882,0.7881,-2.3739\}.
$$

\section*{Acknowledgments} JFC and CLL acknowledge partial
financial support from DGICYT under project
PB--93.0582

\end{document}